\documentstyle[psfig]{aipproc}

\def\la{\mathrel{\hbox{\rlap{\hbox{\lower4pt\hbox{$\sim$}}}\hbox{$<$}}}}
\def\ga{\mathrel{\hbox{\rlap{\hbox{\lower4pt\hbox{$\sim$}}}\hbox{$>$}}}}

\begin{document}
\title{Construction and Preliminary Application of the Variability $\rightarrow$ 
Luminosity Estimator}

\author{Daniel E. Reichart$^{1,2}$ and Donald Q. Lamb$^3$} 
\address{$^1$Department of Astronomy, California Institute of Technology,
Mail Code 105-24, 1201 East California Boulevard, Pasadena, CA 91125\\
$^2$Hubble Fellow\\
$^3$Department of Astronomy \& Astrophysics, University of Chicago, 5640 South 
Ellis Avenue, Chicago IL, 60637}

\maketitle

\begin{abstract}    
We present a possible Cepheid-like luminosity estimator for the long-duration
gamma-ray bursts based on the variability of their light curves.  We also 
present a preliminary application of this luminosity estimator to 907 
long-duration bursts from the BATSE catalog.
\end{abstract}

\section*{Introduction} 

Since gamma-ray bursts (GRBs) were first discovered [1], thousands of bursts
have been detected by a wide variety of instruments, most notably, the Burst and
Transient Source Experiment (BATSE) on the {\it Compton Gamma-Ray Observatory
(CGRO)}, which detected 2704 bursts by the end of {\it CGRO}'s more than 9 year
mission in 2000 June (see, e.g., [2]).  However, the distance scale of the
bursts remained uncertain until 1997, when BeppoSAX began localizing 
long-duration bursts
to a few arcminutes on the sky, and distributing the locations to observers
within hours of the bursts.  This led to the discovery of X-ray [3], optical
[4], and radio [5] afterglows, as well as host galaxies [6].  Subsequent
observations led to the spectroscopic determination of burst redshifts, using
absorption lines in the spectra of the afterglows (see, e.g., [7]), and emission
lines in the spectra of the host galaxies (see, e.g., [8]).  To date, redshifts
have been measured for 13 bursts.

Recently, [9] (see also [10]), [11] (see also [12]), and [13] (see also [14])
have proposed trends between burst luminosity and quantities that can be
measured directly from burst light curves, for the long-duration bursts.  Using
1310 BATSE bursts for which peak fluxes and high resolution light curves were
available, [9] have suggested that simple bursts (bursts dominated by a single,
smooth pulse) are less luminous than complex bursts (bursts consisting of
overlapping pulses); however, see [15].  Using a sample of 7 BATSE bursts for
which spectroscopic redshifts, peak fluxes, and high resolution light curves
were available, [11] have suggested that more luminous bursts have shorter
spectral lags (the interval of time between the peak of the light curve in
different energy bands).  Using the same 7 bursts, [13] have suggested that more
luminous bursts have more variable light curves.  These trends between
luminosity and quantities that can be measured directly from light curves raise
the exciting possibility that luminosities, and hence luminosity distances,
might be inferred for the long-duration bursts from their light curves alone.

\begin{figure}[t]
\centerline{\psfig{file=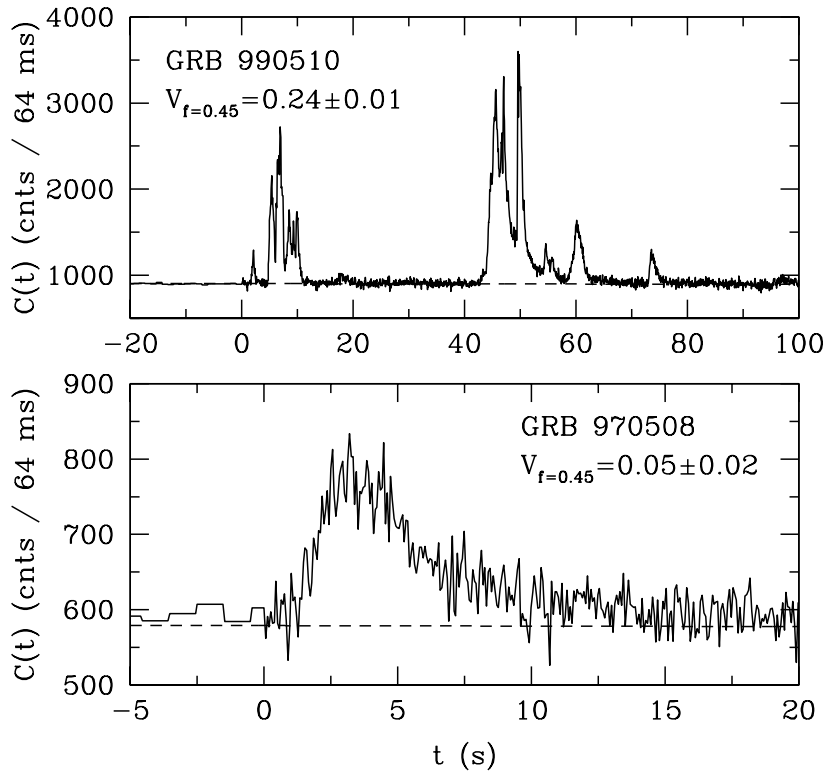,width=4.5truein,clip=}}
\caption{The $> 25$ keV light curves of the most (GRB 990510)
and least (GRB 970508) variable cosmological BATSE bursts in our sample.  In the
case of GRB 990510 ($z = 1.619$), we find that $V = 0.24 \pm 0.01$.  In
the case of GRB 970508 ($z = 0.835$), we find that $V = 0.05 \pm
0.02$.}
\end{figure}

\begin{figure}[t]
\centerline{\psfig{file=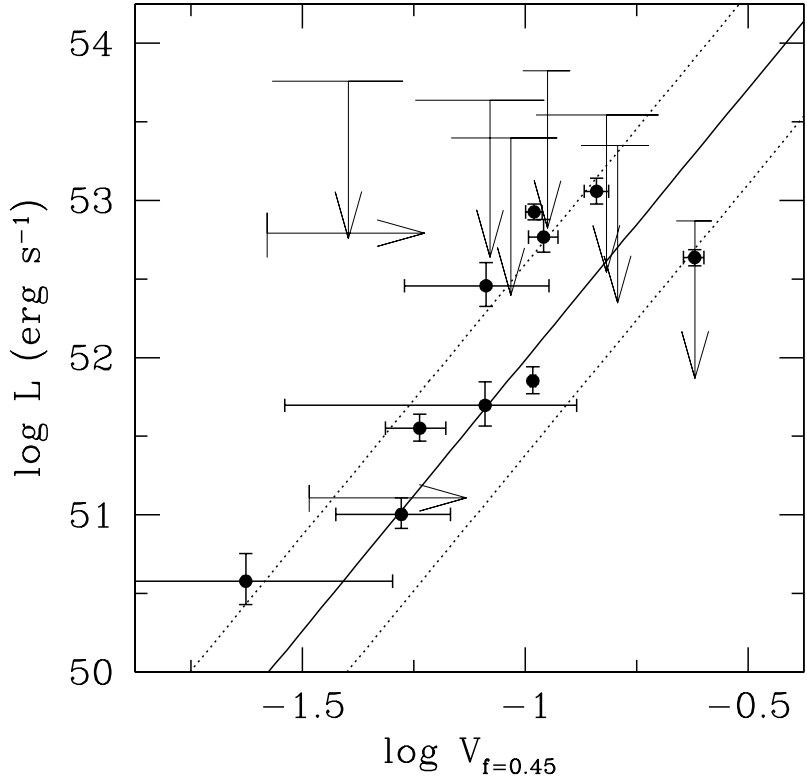,width=4.5truein,clip=}}
\caption{The variabilities $V$ and
isotropic-equivalent peak photon luminosities $L$ between source-frame energies
100 and 1000 keV (see [15]) of the bursts in our sample,
excluding GRB 980425.  The solid and dotted lines mark the center and 1 $\sigma$
widths of the best-fit model distribution of these bursts in the
$\log{L}$-$\log{V}$ plane.}
\end{figure}

In this paper (see also [15,16]), we present a possible luminosity estimator for
the long-duration bursts, the construction of which was motivated by the work of
[14] and [13].  We term the luminosity estimator ``Cepheid-like'' in that it can
be used to infer luminosities and luminosity distances for the long-duration
bursts from the variabilities of their light curves alone.  We also present a 
preliminary application 
of this luminosity estimator to 907 long-duration bursts from the BATSE catalog.

We discuss the construction of our measure $V$ of the variability of a burst 
light curve \S 2.  In
\S 3, we discuss our expansion of the original [14] sample of 7 bursts to
include a total of 20 bursts, including 13 BATSE bursts, 5 {\it Wind}/KONUS
bursts, 1 {\it Ulysses}/GRB burst, and 1 NEAR/XGRS burst.  Also in \S 3, we
discuss the construction of our luminosity estimator.  We present our
preliminary application of this luminosity estimator in \S 4.

\begin{figure}[t]
\centerline{\psfig{file=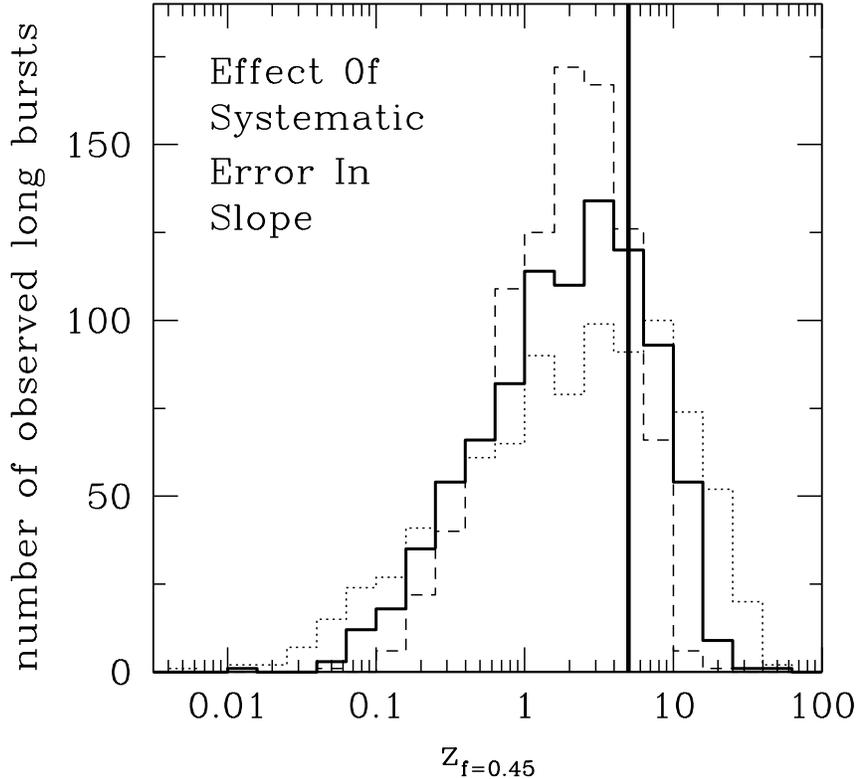,width=4.5truein,clip=}}
\caption{Solid histogram:  The distribution of variability redshifts as 
determined using the best-fit luminosity estimator (Figure 2).  Dotted and 
dashed histograms:  The effect of varying the fitted slope of the luminosity 
estimator by $\pm$ its 1 $\sigma$ statistical uncertainty.  The vertical line 
marks $z = 5$.}
\end{figure}

\section*{The Variability Measure}

Qualitatively, $V$ is computed by taking the difference of the light curve and a
smoothed version of the light curve, squaring this difference, summing the
squared difference over time intervals, and appropriately normalizing the
result.  We rigorously construct $V$ in [15].  We require it to have the
following properties:  (1) we define it in terms of physical, source-frame
quantities, as opposed to measured, observer-frame quantities; (2) when
converted to observer-frame quantities, all strong dependences on redshift and
other difficult or impossible to measure quantities cancel out; (3) it is not
biased by instrumental binning of the light curve, despite cosmological time
dilation and the narrowing of the light curve's temporal substructure at higher
energies [17]; (4) it is not biased by Poisson noise, and consequently can be
applied to faint bursts; and (5) it is robust; i.e., similar light curves always
yield similar variabilities.  Also in [15], we derive an expression for the
statistical uncertainty in a light curve's measured variability, and we describe
how we combine variability measurements of light curves acquired in different
energy bands into a single measurement of a burst's variability.  We plot the $>
25$ keV light curves of the most and least variable cosmological BATSE bursts in
our sample in Figure 1.

\begin{figure}[t]
\centerline{\psfig{file=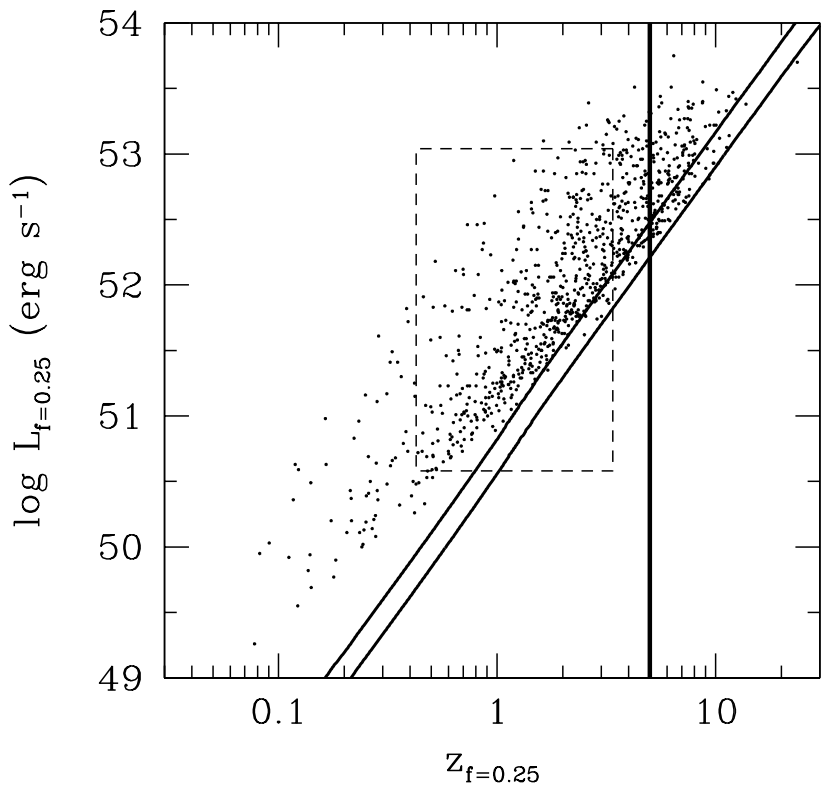,width=4.5truein,clip=}}
\caption{The joint redshift and luminosity distribution of the qualitatively 
acceptable redshift distribution of Figure 3 (the dashed histogram).  The solid 
curves mark the 90\% and 10\% detection thresholds of BATSE.  The dashed box 
marks the redshift and luminosity ranges of the bursts that were used to 
construct the luminosity estimator, everything outside of this box is 
technically an extrapolation.  The vertical line marks $z = 5$.}
\end{figure}

\section*{The Luminosity Estimator}

We list our sample of 20 bursts in Table 1 of [15]; it consists of every burst
for which redshift information is currently available.  Spectroscopic redshifts,
peak fluxes, and high resolution light curves are available for 11 of these
bursts; partial information is available for the remaining 9 bursts. We
rigorously construct the luminosity estimator in [15], applying the Bayesian
inference formalism developed by [18].  We plot the data and best-fit model of
the distribution of these data in the $\log{L}$-$\log{V}$ plane in Figure 2.

\section*{Preliminary Application}

We apply the best-fit luminosity estimator (Figure 2) to 1667 BATSE bursts,
which are all of the BATSE bursts for which the necessary information was
available as of the summer of 2000.  To remove the short-duration bursts, we
conservatively cut the bursts with durations of $T_{90} < 10$ sec from the
sample, reducing the number to 907.  We plot the distribution of variability 
redshifts in Figure 3.  A rigorous analysis of how statistical and systematic 
errors affect this distribution will be presented elsewhere, but the largest 
effect is the statistical uncertainty in the fitted slope of the luminosity 
estimator.  We plot how reasonable variations of this slope affect the 
distribution also in Figure 3.  Although the original distribution has too many 
low-$z$ and high-$z$ bursts for comfort, reasonable variations appear to yield 
at least qualitatively acceptable distributions (see Figure 5 of [19]).

We plot the joint redshift and luminosity distribution of the qualitatively 
acceptable redshift distribution of Figure 3 (dashed histogram) in Figure 4.  If 
systematic effects can be ruled out near BATSE's detection threshold, this 
distribution suggests that the luminosity distribution of the bursts is 
evolving, in which case no more than about 15\% of bursts have redshifts greater 
than $z = 5$, in contrast to the results of [13] and [20].

\end{document}